\documentstyle[preprint,revtex,eqsecnum]{aps}

\newcommand{\be}{\begin{equation}}
\newcommand{\ee}{\end{equation}}
\newcommand{\beqn}{\begin{eqnarray}}
\newcommand{\eeqn}{\end{eqnarray}}
\newcommand{\bma}{\begin{mathletters}}
\newcommand{\ema}{\end{mathletters}}

\begin{document}
\begin{title}
Chiral Vertex Operators in Off-Conformal Theory:\\
The Sine-Gordon Example\\
\end{title}
\author{Shau-Jin Chang}
\begin{instit}
Department of Physics\\
University of Illinois at Urbana-Champaign\\
1110 West Green Street \\
Urbana, Illinois  61801-3080 USA\\
\end{instit}
\author{R. Rajaraman\cite{byline}}
\begin{instit}
Department of Physics\\
University of Illinois at Urbana-Champaign\\
1110 West Green Street\\
Urbana, Illinois  61801-3080 USA\\
and\\
Indian Institute of Science\\
Bangalore 560012, India
\end{instit}
\vspace{60 mm}

\small\hspace{2.0 in}Typed by:  Sandy Chancey/Char Cleek-Agans

\newpage

\begin{abstract}

\begin{center}
{Abstract}
\end{center}

We study chiral vertex operators in the sine-Gordon [SG] theory, viewed as an
off-conformal system.  We find that these operators, which would have been
primary fields in the conformal limit, have interesting and, in some ways,
unexpected properties in the SG model.  Some of them continue to have scale-
invariant dynamics even in the presence of the non-conformal cosine
interaction.  For instance, it is shown that the Mandelstam operator for the
bosonic representation of the Fermi field does {\it not} develop a mass
term in the SG theory, contrary to what the real Fermi field in the massive
Thirring model is expected to do.  It is also shown that in the presence of
the non-conformal interactions, some vertex operators have unique Lorentz
spins, while others do not.

\end{abstract}
\newpage

\section{Introduction}

In this paper, we derive some interesting properties of vertex functions in
the quantum sine-Gordon [SG] model.  This model has been under study for a
long time.  Its several remarkable properties, such as its integrability and
exact solubility, the presence of soliton solutions and topological sectors,
the availability of its exact S-matrix etc, have been known since the
seventies.  In this paper, however, our interest derives from a somewhat
different feature of the SG model, namely, that it can be viewed as a very
attractive example of an off-conformal system.  In recent years there has been
considerable interest in studying off-critical (off-conformal) theories by
techniques of conformal perturbation theory.  In particular, Zamalodchikov
showed that for certain off-conformal theories, new conserved currents could
be found through such methods.$^{1}$  Subsequently Bernard and LeClair applied
these methods to the SG system viewed as an off-conformal theory.$^{2}$  They
derived new nonlocal currents for this system which obey a very interesting
algebraic and braiding properties, powerful enough to lead to its exact S-
matrix.  The fermionic version of these currents was also identified.$^3$

Our interest in this area was triggered by the Bernard-LeClair work.  They had
identified their new conserved currents for the SG model within the conformal
perturbation theory (CPT) framework.  We were interested in seeing whether
such results obtained in the CPT framework also hold, literally in the same
form, under conventional canonical quantization in familiar Minkowskian time.

We showed in an earlier paper that in fact they don't, and require
modification.$^{4}$  Specifically, we found exact expressions for the SG
nonlocal currents which were similar to, but not precisely the same as those
quoted by Bernard and LeClair.  The reason for this discrepancy is presumably
because in CPT, fields are schematically denoted by their deep ultra-violet
behavior, rather than their actual behavior in all of space-time in the off-
conformal system.  We also found that Zamolodchikov's argument for the strict
absence of higher order corrections to these nonlocal currents manifests
itself in the canonical framework through the fact that certain vertex
operators commute with the interaction.

Such results motivated us to study more systematically in this paper, the
properties of general chiral vertex functions in the SG system, under
canonical quantization.  Recall that in the absence of the cosine interaction,
the SG system is a free massless scalar field, which is the prototype example
of a conformally invariant field theory, with a pair of chiral fields which
are respectively left and right moving.  Chiral vertex operators, which are
exponentials of these chiral fields, play a major role in that conformal
theory by providing an infinite number of primary fields, each characterized
by a different pair of Lorentz spin and scale dimensions.  We wanted to show
how the behavior of these vertex operators under time-evolution, Lorentz boost
and scaling, is affected by the presence of the non-conformal $-
\lambda(\cos\beta\phi)$ interaction in the SG theory.

We find several interesting results.  Chiefly,\\
(i) There exist families of local vertex operators which commute with the
interaction potential of the SG theory, for $\beta^2<1$.  They therefore
evolve in time in a scale invariant fashion, as if this non-conformal
interaction were simply not there.\\
(ii) A particularly interesting example of this family is the Mandelstam
construction of the Fermi field as part of the bosonization programm.$^{5,6}$
This Mandelstam operator in fact turns out not to contain a mass term in its
equation of motion for $\beta^2<1$, in contrast to what equivalence to the
massive Thirring model would lead us to expect.  We find that it is only when
$\beta = 1$, which corresponds to free fermions, that the Mandelstam operator's
dynamics has a finite mass term.\\
(iii) We find that, the Lorentz spin of vertex operators is affected by the
presence of the interaction term.  Unlike in the conformal limit, the vertex
operators here do not carry a unique Lorentz spin.  Special subsets of them
which do, are identified. \\
(iv) Surprisingly, despite the presence of the scale-breaking interaction, the
vertex operators continue to be eigenfunctions of the scaling operator at any
given instant.

Even though all our results are obtained only for the SG theory, we believe
that our analysis carries useful lessons about off-conformal theories in
general.

This paper is structured as follows:  In sec. 2, we gather together basic
equal-time commutators of the chiral fields, and define chiral vertex
operators, keeping track of potential ultraviolet and infra-red divergences of
this ($1 + 1$) dimensional theory.  Some inputs for this section are detailed
in Appendix A.  In sec. 3, we show that certain classes of vertex operators
commute with the interaction, and discuss the significance of this fact to the
nonlocal conserved currents obtained in ref. 4.  In sec. 4, we point out how
for the above classes of vertex functions, which include the Mandelstam
operator, their dynamics is scale invariant.  In sec. 5, the scaling and
Lorentz boost behavior of vertex operators is discussed.

\section{Preliminaries}

We will treat the sine-Gordon [SG] system using the canonical equal-time
algebra of chiral fields and their vertex functions.  Much of conformal field
theoretic literature deals with such fields using operator product expansions
in terms of complex Euclidean coordinates.  Equal-time Minkowskian commutators
for chiral fields are less familiar.  Therefore for completeness we will
gather together in this section a few basic equal-time formulae needed for our
calculations in the subsequent sections.

Consider a relativistic scalar field $\phi(x,t)$ in $1 + 1$ dimensions governed
by a Lagrangian density of the form

\begin{equation}
{\cal L} = \frac{1}{8\pi} \dot{\phi}^2 - \frac{1}{8\pi}
\left(\frac{\partial \phi}{\partial x}\right)^2 - V(\phi)
\end{equation}

\noindent
where dot refers to time-derivative.  For the SG theory, the classical
potential is

\begin{equation}
V(\phi) = - \lambda \cos \beta \phi .
\end{equation}

Let us define ``chiral'' fields $\rho$ and $\overline{\rho}$ by
\bma
\be
\rho(x,t)  \equiv  \frac{1}{2}\left[\phi(x,t)
            +\int_{-\infty}^{x}\dot{\phi}(y,t)dy\right]
\ee
\be
\overline{\rho}(x,t)  \equiv
        \frac{1}{2}\left[\phi(x,t)-\int_{-\infty}^{x}\dot {\phi} (y,t)dy\right]
\ee
\ema
Clearly $\phi = \rho + \overline{\rho}$.  It is only in the absence of the
interaction $V(\phi)$ that $\rho(\overline{\rho})$ would be chiral in the
sense of being purely left (right) moving.  But we will retain the definition
in (2.3) of $\rho$ and $\overline{\rho}$ even for the non-conformal case
$(V(\phi)\ne 0)$, and we will continue to call them chiral fields for
simplicity.

In the quantized theory, the fields $\phi$ and $\dot{\phi}$ obey the
standard canonical equal-time commutation rules
\bma
\be
[\phi(x), \dot{\phi}(y)]_{t} = 4\pi i \delta(x-y) \, ,
\ee
\be
[\phi(x), \phi(y)]_{t} = [\dot{\phi}(x), \dot{\phi}(y)]_t=0 \, .
\ee
\ema
Henceforth, all our operators, their products and commutators will be at some
common time $t$.  With this understanding, the symbol $t$ will not always be
displayed.  The canonical commutators in (2.4) form the basis of all our
calculations.  It must be remembered that they hold exactly regardless of
whether the interaction $V(\phi)$ is present or not, or what form $V(\phi)$ has
(as long as this interaction does not involve $\dot{\phi}$).  We stress this
obvious point to contrast our approach with certain conformal perturbation
theoretical methods, which perturbatively employ expressions for time-ordered
operator products that really hold only in the conformal $(V(\phi) = 0)$
limit.

Given (2.4) the equal-time commutators of the chiral fields are
\[
[\rho (x), \rho (y)] = [\overline{\rho} (y), \overline{\rho} (x)] = -i \pi
\epsilon (x-y) \, ,
\]
\be
[\rho (x), \overline{\rho} (y)] = -i \pi \, .
\ee
Next, we separate fields into creation and annihilation parts for purposes of
normal ordering, to render our operators finite in the ultraviolet.  All our
results arise basically due to short-distance singularities, but in this
$(1+1)$ dimensional theory, care has also to be exercised regarding possible
infrared divergences.  Although our final results are in fact free of
infrared divergences, we will keep careful track of infrared behavior.  Let
us normal order the fields $\phi$ and $\dot{\phi}$ (and thereby also $\rho$
and $\overline{\rho}$ through their definitions (2.3)) with respect to a free
field of {\it nonzero} mass $m$.  The resulting equal-time commutators
at some common time $t$ between the creation ($\rho_{-}, \overline{\rho}_{-}$)
and destruction ($\rho_{+}, \overline{\rho}_{+}$) parts are shown in Appendix
A to be
\beqn
\left[\rho_{+}(x), \rho_{-}(y)\right] & = & K_{0}(m| x-y|)+\frac{1}{2}M
(m|x-y|) - i\frac{\pi}{2}\epsilon(x-y)\, \nonumber\\
\left[\rho_{+}(x), \overline{\rho}_{-}(y)\right] & = & -\frac{1}{2}
 M(m|x-y|) - i\frac{\pi}{2} \, \nonumber\\
\left[\overline{\rho}_{+}(x), \overline{\rho}_{-}(y)\right] & = &
K_{0}(m|x-y|) + \frac{1}{2}M(m|x-y|) + i\frac{\pi}{2}\epsilon(x-y) \,
\nonumber\\
\left[\overline{\rho}_{+}(x), \rho_{-}(y)\right] & = &
-\frac{1}{2}M(m| x-y |) + i\frac{\pi}{2} \,
\eeqn
where $K_{0}(z)$ is a Bessel function.  As shown in Appendix A, the
function $M(z)$ which obeys
\be
\frac{d^{2}M(z)}{d z^2} = - K_{0}(z) \,
\ee
actually contains an infrared divergent, but $z$-independent integration
constant.  That will affect us only to the extent of having to insert an
infrared renormalization constant into our definiton of vertex operators (see
below). Aside from this, for the bulk of our work, we will need only the
following equal-time commutators, obtained from eq (2.6) using
$\phi=\rho+\overline{\rho}$,
\[
[\rho_{+}(x), \phi_{-}(y)] = K_{0}(m | x-y | ) - i \pi \theta (x-y)
= -\ln (i \overline{m} (x-y-i\epsilon)) - \frac{i\pi}{2}
+ O ((x-y)^{2}) \,
\]
\[
[\rho_{-}(x), \phi_{+}(y)] = -K_{0} (m | x-y |) - i\pi \theta (x-y)
= \ln (-i\overline{m} (x-y+i\epsilon)) - \frac{i\pi}{2}
+ O ( (x-y)^{2})\,
\]
\[
[\overline{\rho}_{+}(x), \phi_{-}(y)] = K_{0} (m | x-y |) + i\pi
\theta (x-y) = -\ln (-i\overline{m}(x-y+i\epsilon)) + \frac{i\pi}{2}
+ O((x-y)^{2}) \,
\]
\be
[\overline{\rho}_{-}(x), \phi_{+}(y)] = -K_{0} (m | x-y |) + i \pi
\theta (x-y) = \ln (i\overline{m} (x-y-i\epsilon)) + \frac{i\pi}{2}
+ O ((x-y)^{2})
\ee
where $\epsilon \rightarrow 0$ represents our short-distance (ultraviolet)
regularization.  We see that these commutators are free of the infrared
divergent function $M(z)$.  The middle expression in each equation in (2.8) is
exact, but we have also shown its behavior in the small (x-y) limit for later
use.  Here $\overline{m} \equiv \frac{1}{2} me^{\gamma}$ where $\gamma$ is
Euler's constant.

We will primarily be interested in vertex operators $W_{a,b}(x)$ defined by
\be
W_{a,b}(x) \equiv D_{a,b}:e^{ia\rho(x) + ib\overline{\rho}(x)}: \equiv
D_{a,b}e^{ia\rho_{-}(x) + ib\overline{\rho}_{-}(x)} e^{ia\rho_{+}(x) +
ib\rho_{+}(x)}
\ee
where $D_{a,b} = \exp [-(a-b)^{2} (m-\mu)/4\mu]$ is a constant with the
infrared cutoff $\mu \rightarrow 0$.  This factor is inserted because of the
infrared divergent constant in the commutators (2.6).  Notice that $D_{a,b}$
is unity when $a=b$, i.e. when only the combination $\phi = \rho +
\overline{\rho}$ is involved. The symbol :: refers to normal ordering with
respect to a nonzero mass $m$ as outlined above. This $W_{a,b}$ is free of
ultraviolet and infrared singularities.

The regularized Hamiltonian of the quantized $SG$ theory can be written as
$H = H_{0} + H_{I}$ where
\bma
\be
H_{0} = \frac{1}{4\pi} \int dy \left( : \left(\frac{\partial \rho}{\partial y}
\right)^{2} + \left(\frac{\partial \overline{\rho}}{\partial y}\right)^{2}:
\right)
\ee
and
\be
H_{I} = -\lambda_m \int : \cos \beta \phi (y): dy = -\frac{\lambda_m}{2}
\int \left(W_{\beta,\beta}(y) + W_{-\beta,-\beta}(y)\right)dy \ .
\ee
\ema
Here $\lambda_m$ is the renormalized coupling constant.  It is important to
remember for our later use that $: \cos \beta \phi:$ is divergence-free,
$\lambda_m$ has to be taken {\it finite} so that the energies and masses
in the vacuum sector of the $SG$ model may be finite.  The constant $\beta$
need not be renormalized and is taken to be in the range $0 < \beta^{2} < 2$
as required by Coleman$^{5}$.

Two properties of $W_{a,b}$, obtainable from the commutators (2.5), (2.6) and
(2.8), and useful for our calculations are
\beqn
\left[W_{a,b}(x), \left(\frac{\partial \rho}{\partial y} \right)^{2} \right.
& + & \left. \left(\frac{\partial \overline{\rho}}{\partial y}
\right)^{2}\right] \nonumber\\
& = & 4\pi i\left[:\left(a \frac{\partial \rho}{\partial x} - b \frac{\partial
\overline{\rho}}{\partial x} \right) W_{a,b}:\delta(x-y)+\frac{a^{2}-
b^{2}}{2}W_{a,b} \partial_{x} \delta(x-y)\right] \,
\eeqn
and the equal-time operator product, obtained by using (2.8)
\beqn
W_{a,b}(x) W_{c,c}(y) & = & D_{a,b}:e^{ia\rho(x)+ib\overline{\rho}+ic\phi(y)}:
\nonumber \\
& \times & \exp \left[-(ac + bc) K_{0}(m | x-y | ) + i\pi(ac-bc)
  \theta(x-y)\right] \ .
\eeqn
Note that the infrared factor $D_{a,b}$, which arises because this same factor
is present in $W_{a,b}$ (see (2.9)), once again precisely offsets the infrared
divergence of the vertex function on the r.h.s.  The only singularity in
(2.12) is at x=y due to the $\theta$ function.

\section{Vertex Operators}

In this section we present some very interesting properties of vertex
operators $W_{a,b}(x)$ in the SG theory.  These properties turn out to be
responsible, among other things for (a) the existence of an exactly conserved
nonlocal current in the SG theory with terms only up to $O(\lambda_m)$ and
(b) the absence of a ``mass term'' in the dynamics of the Mandelstam fermion
operator even in the presence of the non-conformal $(-\lambda_{m}:\cos \beta
\phi:)$ term in the Hamiltonian.

We will consider only those $W_{a,b}$ which satisfy
\be
(a-b)\beta =n
\ee
where $n$ is an integer, positive, negative or zero.  The reason for placing
this restriction comes from the topological index of $W_{a,b}$, evaluated in
ref. (2).  As is well known, all finite energy configurations in the $SG$
theory must carry an integer value for the topological quantum number
\be
T \equiv \frac{\beta}{2\pi} \int dx \frac{\partial \phi}{\partial x} =
\frac{\beta}{2 \pi} \int dx \left(\frac{\partial \rho}{\partial x} +
\frac{\partial \overline{\rho}}{\partial x}\right) \, .
\ee
{}From the commutation rule (2.5), we see that
\be
\left[ \frac{\partial \rho}{\partial x} +
\frac{\partial \overline{\rho}}{\partial x}, W_{a,b}(y) \right]
= \pi (a-b) \frac{\partial}{\partial x} \epsilon(x-y) W_{a,b}(y) \ .
\ee
Hence,
\be
\left[T, W_{a,b}(y) \right] = (a-b) \beta W_{a,b}(y) \ .
\ee
Thus the requirement (3.1) ensures that the $W_{a,b}$ have integer valued
topological quantum numbers.  Under the equivalence to the Thirring
model,$^{5}$ one expects the $W_{a,b}$ satisfying (3.1) to have a nonzero
matrix element between the vacuum sector and the $n$-fermion sector.

Take any such $W_{a,b}(x)$ satisfying $(a-b) \beta = n$, where to be specific
let $n > 0$.  The $n \leq 0$ cases are treated similarly (see
below). Consider the commutator of such a $W_{a,b}$ with the interaction
Hamiltonian $H_{I} = -\frac{\lambda_{m}}{2} \int dy
\left(W_{\beta,\beta}(y) + W_{-\beta,-\beta}(y) \right)$.  By translation
invariance it is sufficient to consider $W_{a,b}(x)$ at $x=0$.

We have, using (2.12),
\beqn
\left[W_{a,b}(0), W_{\beta,\beta}(y) \right] = & C_{+}(y) \left\{ \left(\exp
[-(a+b)\beta K_{0}(m \left| -y \right|)+i\pi(a-b)\beta\theta(-y)\right]\right.
\nonumber \\
& \left. -\exp [-(a+b)\beta K_{0}(m | y | ) + i\pi(a-b) \beta \theta(y) ]
\right\}
\eeqn
where
\be
C_{\pm}(y) \equiv D_{a,b}:e^{ia\rho(0) + ib \overline{\rho}(0)
\pm i\beta \phi(y)}:
\ee
are nonsingular operators.

We see that since $(a-b) \beta = n$, the r.h.s. of (3.5) vanishes for $y \neq
0$. Hence, if this commutator were to be integrated over $y$, the only
contribution will come from possible singularities at $y=0$.  Therefore, we
can use the short distance behavior of these commutators, given in (2.8) to
write
\beqn
\left[W_{a,b}(0), -\frac{\lambda_m}{2} \int dy
W_{\beta,\beta}(y)\right] & \nonumber \\
= -\frac{\lambda_m}{2} \lim_{\epsilon \rightarrow 0}
\int dy \, C_{+}(y) & \{(\overline{m} (-y-i\epsilon))^{a\beta}
(\overline{m}(-y +i\epsilon))^{b\beta} e^{i\pi(a-b)\beta} \nonumber \\
& -(\overline{m}(-y+i\epsilon))^{a\beta}(\overline{m}(-y-i\epsilon))^{b\beta}
e^{-i\pi(a-b) \beta} \} \nonumber \\
= -\frac{\lambda_m}{2} (\overline{m})^{(a+b)\beta}
\lim_{\epsilon \rightarrow 0} \int dy & C_{+}(y)(y^{2}+\epsilon^{2})^{b\beta}
\{ (y+i\epsilon)^{n}-(y-i\epsilon)^{n} \} \ .
\eeqn
Similarly
\beqn
\left[W_{a,b}(0), -\frac{\lambda_m}{2} \int dy
W_{-\beta,- \beta}(y) \right] & \nonumber \\
= -\frac{\lambda_m}{2}(\overline{m})^{-(a+b)\beta} \lim_{\epsilon
\rightarrow 0} \int dy & C_{-}(y)(y^{2}+\epsilon^{2})^{-a\beta}
\{(y+i\epsilon)^{n} - (y - i\epsilon)^{n}\} \ .
\eeqn
The operators $C_{\pm}(y)$ defined in (3.6) are properly regularized and
singularity-free.  The integrands in (3.7-3.8) vanish as $\epsilon \rightarrow
0$ for any $y \neq 0$.  Hence, the $y$-integrals will vanish unless there is a
sufficiently strong singularity at $y = 0$.  For any given positive integer
$n$, such singularities in the commutator can arise only for sufficiently
negative $b$ in (3.7) or positive $a$ in (3.8).  Very roughly, one would
estimate that the commutator (3.7) behaves like $(y^{2}+\epsilon^{2})^{(b\beta
+ n)}C_{+}(y) \partial_{y}^{n-1} \delta(y)$ and that its integral over $y$
would behave as $\epsilon^{2\beta b+n+1}$ with $\epsilon \rightarrow 0$.  This
expectation is made more precise in Appendix B where we evaluate, for any
positive integer $n$, any real $p$ and any nonsingular function $C(y)$, the
integral
\be
I\equiv \lim_{\epsilon \rightarrow 0} \int dy
(y^{2}+\epsilon^{2})^{p}\left((y+i\epsilon)^{n}-(y-i\epsilon)^{n}\right)C(y)
\ .
\ee
It is shown here that this integral vanishes if $2p+n+1>0$ when $n$ is odd and
if $2p+n+2>0$ when $n$ is even.  Let us apply this to eq. (3.7) and (3.8). The
term in (3.7) vanishes if $2p\beta +n+1>0$ for $n$ odd and $2b\beta + n+2>0$
for $n$ even. Along with (3.1), this condition becomes
\bma
\be
a\beta >\frac{n-1}{2}\qquad\qquad \mbox{for $n$ odd}
\ee
and
\be
a\beta >\frac{n-2}{2} \qquad\qquad \mbox{for $n$ even} \ .
\ee
\ema
Similarly, (3.8) vanishes if $-2a\beta + n + 1>0$ for $n$ odd and $-2a\beta +
n + 2>0$ for $n$ even, which amounts to
\bma
\be
a\beta <\frac{n+1}{2} \qquad\qquad\mbox{for $n$ odd}
\ee
and
\be
a\beta < \frac{n+2}{2} \qquad\qquad\mbox{for $n$ even} \ .
\ee
\ema
By adding (3.7) and (3.8), we therefore see (since the renormalized coupling
$\lambda_m$ is finite) that
\be
[W_{a,b}(0),H_{I}] = [W_{a,b}(0), -\lambda_{m} \int dy : \cos \beta
\phi(y):] = 0
\ee
for any finite $\lambda_{m}$, provided
\bma
\be
\frac{n-1}{2}<a\beta <\frac{n+1}{2} \qquad\qquad\mbox{for $n$ odd}
\ee
or
\be
\frac{n-2}{2}<a\beta <\frac{n+2}{2} \qquad\qquad\mbox{for $n$ even}
\ee
\ema
where we recall that $b$ is fixed for any given $a$ and $n$ by (3.1).

We have analyzed the case when the integer $n>0$.  Similar results also hold
for $n \leq 0$.  When $n=0, a=b$ and $W_{a,b} = : e^{ia\phi}:$ and
clearly this commutes with $:\cos \beta \phi(y):$ for all $a$.  The result for
the $n<0$ case is treated in the same way as the $n>0$ case, with the roles of
$a$ and $b$ interchanged. The condition (3.13) continues to hold unchanged
even for $n<0$.  Hence, for any integer $n, W_{a,b}$ with $a$ and $b$
satisfying (3.1) and (3.13) commutes with the interaction Hamiltonian $H_{I}$.

It should be emphasized that these results cannot be circumvented by a further
ultraviolet renormalization of $\lambda_m$ as $\epsilon \rightarrow 0$. The
operator $: \cos \beta \phi:$ has been rendered finite in the vacuum sector by
normal ordering. Hence $\lambda_{m}$ {\it has} to be taken finite in
order to get $a$ finite energies into the vacuum sector.  There is no further
freedom to infinitely renormalize $\lambda_m$ to compensate for the zeroes of
(3.7) and (3.8) as $\epsilon \rightarrow 0$.

The vanishing of (3.12) holds when the inequalities (3.13) are satisfied along
with (3.1).  A marginal situation arises when these conditions become
equalities.  For such cases, we apply equations (B.5b) and (B.6b) in Appendix
B to the integrals in (3.7-3.8).  We see that when
\bma
\be
(a-b)\beta = n, \quad a \beta = \frac{n\pm 1}{2} \qquad\qquad\mbox{(odd $n$)}
\ee
or
\be
(a-b)\beta = n, \quad a \beta = \frac{n\pm 2}{2} \qquad\qquad\mbox{(even $n$)}
\ee
\ema
then the commutator $[W_{a,b}(x), H_{I}]$ is finite and generically nonzero,
as $\epsilon \rightarrow 0$.  When the values of (a,b) fall outside the above
ranges, i.e. when $a \beta > \frac{n + 1}{2}$, or $a \beta < \frac{n-1}{2}$
(for $n$ odd) and $\alpha \beta > \frac{n+2}{2}$ or $\alpha \beta < \frac{n-
2}{2}$ (for $n$ even), this commutator diverges.

The vanishing of the commutator (3.12) for some ranges of $a$ and $b$ has
several interesting consequences.  One application arises in the derivation of
exact nonlocal currents for the SG model$^{2,4}$.  Since the derivation of
these currents using canonical equal-time methods has already been given
briefly in ref. (4) and in Appendix A, we will not go into all the details
here, other than to point out how our result (3.12) is crucial for the success
of the Zamoldchikov's argument$^{1}$ (3.12) involved there.  This argument
says that in such currents obtained for off-conformal theories with relevant
perturbations (of which the SG model is an example), terms of second and
higher order in the perturbation will be strictly absent.  In the papers of
Zamoldchikov$^{1}$ and Bernard and Leclair,$^{2}$ this argument is based on
scale-weight counting within the conformal perturbation theoretic framework.
The corresponding manifestation of this phenomenon in canonical equal-time
quantization relies on our (3.12).  To see this, recall$^{2,4}$ that to obtain
this nonlocal current in the SG theory, one starts with the operator
$W_{\frac{2}{\beta},0} = D_{\frac{2}{\beta}},0
:e^{\frac{2i}{\beta}\rho(x,t)}:$.  In the absence of the interaction $H_{I} =
-\lambda_m: \cos \beta \phi:$, $\rho(x,t)$ is purely left moving, so that, to
order $(\lambda_m)^{0}$,
\be
\partial_{-}W_{\frac{2}{\beta},0}(xt) \equiv (\partial_{t}-\partial_{x})
W_{\frac{2}{\beta},0}(x,t) = 0 \ .
\ee
When $H_{I}$ is turned on, eq. (3.15) is modified to
\be
\partial_{-}W_{\frac{2}{\beta},0} = [W_{\frac{2}{\beta},0}(x,t), H_{I}] \ .
\ee
Notice that $W_{\frac{2}{\beta},0}$ corresponds to $a = \frac{2}{\beta}, b =
0$ and $n = \beta (a-b) = 2$.  Thus it satisfies (3.1) and is a legitimate
operator with topological number 2.  Further, since it satisfies the equality
condition (3.14b), its commutator with $H_{I}$ will be finite and nonzero.
Indeed this is what the explicit calculation in ref. (4) shows, where after
some algebra, the eq. (3.16) is put in the form (see eq. (3.8) and (3.9) of
ref. 4),
\be
\partial_{-}W_{\frac{2}{\beta},0}(x) = (B_{1} \partial_{+} + B_{2} \partial
_{-}) W_{\frac{2}{\beta} -\beta, -\beta}(x) + B_{3} [W_{\frac{2}{\beta} -
\beta, -\beta}(x), H_{I}]
\ee
where $B_{1}, B_{2}$ and $B_{3}$ are constants of order $\lambda_m$.  Now, the
last term in (3.17) is $O(\lambda_m)$ since $B_{3}$ and $H_{I}$ are each of
order $\lambda_m$.  However, consider the operator $W_{\frac{2}{\beta} -\beta,
-\beta}$.  It corresponds to $a = \frac{2}{\beta} -\beta$ and $b = -\beta$, so
that $(a-b) \beta = 2$, once again corresponding to topological number 2.  But
it obeys the inequality (3.13b), since
\be
\frac{n-2}{2} (= 0) < a \beta (= 2 - \beta^{2}) < \frac{n + 2}{2}(= 2)
\ee
for all $0 < \beta^{2} < 2$ which, according to Coleman$^{5}$, is the full
range of $\beta$ for which the SG theory exists.  Consequently, the commutator
in the last term of (3.17) vanishes.  The remaining terms yield an exactly
conserved current containing only terms up to $O(\lambda_m)$, and the
Zamoldchikov conjucture holds in this canonical quantization framework as
well.  For more properties of these currents, see ref. (2, 3 and 4), as well
as later sections of this paper.

\section{Dynamics of the Mandelstam Operator}

Another important consequence of the vanishing of the commutator $[W_{a,b}(x),
H_I]$ for a range of $a$ and $b$ is that the time evolution of $W_{a,b}(x)$ is
not affected by the presence of $H_I$.  These $W_{a,b}$ obey a scale invariant
field equation.  A particularly interesting example of such $W_{a,b}$ is the
Mandelstam operator for the bosonic representation of the Fermi field.

Consider the time evolution of the operator $W_{a,b}(x,t)$ with $a$ and $b$
satisfying (3.1) and (3.13).  We have, in the Heisenberg representation,
\be
\frac{\partial}{\partial t} W_{a,b}(x,t)  =  -i[W_{a,b} , H_0 + H_I]
\ee
where the Hamiltonian is given in (2.10).  Since this $W_{a,b}(x)$
commutes with the interaction Hamiltonian, its time evolution is determined
solely by the kinetic term in $H_0$.  We have, using (2.10), (2.11) and
(3.12),
\be
\frac{\partial}{\partial t} W_{a,b}(x,t) =  -i\left[ W_{a,b}(x) ,
\frac{1}{4\pi} \int dy : \left( (\frac{\partial\rho}{\partial y})^2
+(\frac{\partial\overline{\rho}}{\partial y})^2 \right) : \right] \ .
\ee
We can use the commutators in (2.11) to evaluate this as
\be
\frac{\partial}{\partial t} W_{a,b}(x,t) = i  : (a \frac{\partial\rho}
{\partial x} - b \frac{\partial\overline{\rho}}{\partial x}) W_{a,b} (x,t):\ .
\ee
This is a very interesting result.  The SG theory is not a scale invariant
theory because of the interaction term $-\lambda_m:\cos\beta\phi:$.
Nevertheless, those vertex functions $W_{a,b}(x)$ which satisfy (3.13) and
(3.1), obey an equation of motion (3.17) which {\it is} scale invariant under
$x \rightarrow kx$, $t\rightarrow kt.$  (Recall that $\frac{\partial\rho}
{\partial x}$ and $\frac{\partial\overline{\rho}}{\partial x}$ have scale
dimension of unity; see sec. 4).  This is because these $W_{a,b}$ commute with
the scale-breaking SG interaction.

An especially inetresting example of this is the well-known Mendelstam
construction$^6$ of the bosonic representation for the Fermi field developed
for relating the SG model to the massive Thirring model (MTM).  Recall that it
has the form (eq. 2.8 of ref. 6)
\[
\Psi=\left(\begin{array}{c}
\psi_1\\
\psi_2
\end{array}\right)
\]
with
\bma
\be
\psi_1 = N : \exp \left[-\frac{i}{2} (\frac{1}{\beta} + \beta) \rho +
\frac{i}{2}(\frac{1}{\beta} - \beta) \overline{\rho}\right]
\ee
\be
\psi_2 = -iN : \exp \left[-\frac{i}{2} (\frac{1}{\beta} - \beta) \rho +
\frac{i}{2}(\frac{1}{\beta} + \beta) \overline{\rho}\right]
\ee
\ema
where $N$ is a constant.  (In converting our notation to that of
Mandelstam$^6$ and Coleman$^5$, change $\beta \rightarrow \beta/ \sqrt{4\pi}$
and $\phi \rightarrow \sqrt{4\pi} \phi$).  Thus, both $\psi_1$ and $\psi_2$
belong to the class of $W_{a,b}$ operators, with
\bma
\be
a_1 = -\frac{1}{2} (\frac{1}{\beta} + \beta)  ,\  b_1 = \frac{1}{2}
(\frac{1}{\beta} - \beta)   ,\  (a_1 - b_1) \beta = n_1 = -1
\ee
and
\be
a_2 = -\frac{1}{2} (\frac{1}{\beta} - \beta)  ,\  b_2 = \frac{1}{2}
(\frac{1}{\beta} + \beta)   ,\  (a_2 - b_2) \beta = n_2 = -1 \ .
\ee
\ema
The condition for the vanishing of $[\psi_1(x), H_I]$ and $[\psi_2(x), H_I]$,
transcribed from (3.13a), reduces respectively to
\[
-1 < a_1\beta = -\frac{1}{2} (1+ \beta^2) < 0
\]
and
\be
-1 < a_2\beta = -\frac{1}{2} (1- \beta^2) < 0 \ .
\ee
Both these conditions are satisfied in the regime $0<\beta^{2}<1$.

Therefore, for $\beta^{2}<1$, the time evolution of the operators $\psi_{1}$
and $\psi_{2}$ is governed solely by the unperturbed Hamiltonian $H_0$ in
(2.10a).  We have, as a special case of (4.3),
\be
\frac{\partial \psi_1}{\partial t} = i : \left(\frac{1}{2}(-\beta -
\frac{1}{\beta}) \frac{\partial \rho}{\partial x} - \frac{1}{2}
(\frac{1}{\beta} -\beta) \frac{\partial \overline{\rho}}{\partial x}
\right)\psi_1:
\ee
and
\be
\frac{\partial\psi_2}{\partial t}  =  i  : \left( \frac{1}{2} (\beta -
\frac{1}{\beta}) \frac{\partial\rho}{\partial x} -  \frac{1}{2}(\beta +
\frac{1}{\beta}) \frac{\partial\overline{\rho}}{\partial x} \right) \psi_2:\ .
\ee
The ``mass term'' expected in these equations is absent even though the SG
Hamiltonian contains the interaction $H_{I} = -\lambda_m \cos \beta \phi:$
which is the analogue of the appropriately regularized mass term
$m_{F}Z\overline{\Psi} \Psi$ in the massive Thirring model$^{5,6}$.  To put it
another way the contribution of $H_{I}$ to $\frac{\partial \psi_{1}}{\partial
t}$ in the $SG$ theory behaves (upon inserting the values for $a$ and $b$ from
(4.5) into (3.7-3.8)) as
\be
[\psi_{1,2},H_I]  = \lim_{\epsilon \rightarrow 0} (\overline{m}\epsilon)^
{(1-\beta^2)}(\frac{\lambda_m}{2\overline{m}}) \psi_{2,1} \ .
\ee
This has the correct dimensions and structure of what the ``mass term" should
have, but it vanishes as $\epsilon \rightarrow 0$ when $\beta^{2} < 1$.  We
repeat again that the SG coupling constant $\lambda_m$ multiplying the normal
ordered $:\cos\beta\phi:$ has to be taken finite, to keep energies finite in
the SG vacuum sector.  There is no freedom to further renormalize $\lambda_m$
to offset the vanishing of (4.9) as $\epsilon \rightarrow 0$.  Also since
(4.9) is linear in $\psi_{1,2}$, no $\epsilon$-dependent rescaling of
$\psi_{1,2}$ will restore this mass term. Lastly the constant $\overline{m} =
\frac{1}{2} me^{\gamma}$ is finite and nonzero, where $m$ is our normal
ordering mass.  Thus, we see no way to circumvent the vanishing of this
``mass" term.

Therefore, in the range $\beta^2 < 1$ (which corresponds$^5$ to the Thirring
coupling $g > 0$), the Mandelstam Fermi field does not develop a mass term,
and instead obeys the scale invariant equation of motion (4.7-4.8) that it
would have in the {\it massless} Thirring model. Since the vector current of
the Thirring model corresponds under bosonization rules to $\beta
\epsilon^{\mu \nu}\partial_\nu \phi$, it can be checked that eqs.(4.7-4.8)
indeed correspond to the massless Thirring equation.

When $\beta^2 = 1 (i.e.\ g=0)$, one has an example of the marginal situation
described in eq. (3.14).  Then the commutator $[\psi_{1,2}(x), H_I]$
{\it is} finite and a nonzero mass term appears in $\partial_t \psi_{1,2}$.
This is also evident from eq (4.4).  Thus, it is only when $\beta^2 = 1$,
which corresponds to free massive fermions with zero Thirring coupling, that
the Mandelstam construction $\Psi[\phi]$ has the same dynamics in the SG model
as the Fermi field does in the corresponding Dirac Theory.  When $\beta^2 <
1$, the desired mass term does not develop, and when $\beta^2 > 1$, the
commutator $[\psi_{1,2}, H_I]$ diverges.  Other problems with the Mandelstam
operator when $\beta^2 \neq 1$ had been noted by Schroer and Truong long
ago$^7$.  Our results dealing with its time-evolution are possible related to
those.

It should be pointed out that our derivations give no disagreement with the
bosonization rules developed by Coleman$^5$ which perturbatively relate
correlation functions of fermion bilinears $: \overline{\Psi} \Psi:$ and
$\overline{\Psi}\gamma^{\mu}\Psi$ in the Thirring model to those of
$:\cos{\beta}\phi:$ and $\beta \epsilon^{\mu \nu}\partial_\nu \phi$
respectively in the massless boson theory.  Nor do we disagree with any
equation explicitly given in the Mandelstam paper$^6$. In that paper (see eq.
4.13 of ref. 6) this mass term is not actually evaluated.  It is left in the
form of a commutator of the form $[\psi, M\int Z \overline{\psi}\psi]$.  Our
point is that this commutator vanishes when explicitly evaluated in the SG
system for $\beta^2 < 1 (g > 0)$, and survives as a finite term only when
$\beta^2 = 1 (g = 0)$.  To that extent, the Mandelstam operator does not
reproduce in the SG model the same time-evolution as the Fermi field in the
MTM, except when $\beta^2 = 1$.  For $\beta^2 < 1$, the Mandelstam operator
obeys a scale-invariant massless field equation even though the SG Hamiltonian
carries the scale-breaking $\lambda_m : \cos\beta\phi :$ interaction --- a
somewhat surprising result.

Finally, note that while we have highlighted the case of the Mandelstam
operator $\psi$ because of its obvious importance, that is just one example of
similar behavior by a whole family of vertex operators $W_{a,b}$ satisfying
the conditions (3.1) and (3.13).  For arbitrary $n$, they are generalizations
of the Mandelstam construction corresponding to arbitrary fermion number.
When $a$ satisfies (3.13), they all satisfy scale-invariant field equations.

\section{Transformation Properties of $W_{\lowercase{a,b}}$ and Currents}

In this section, we shall work out the transformation properties of the vertex
operator $W_{a,b}(x,t)$ and currents $J_{\pm}$ under Lorentz and scale
transformations.  We assume that all operators are normal-ordered with respect
to the reference mass $m$. We shall omit its subscript $m$ and denote the
coupling constant simply as $\lambda$.  In Appendix A, we show how $W_{a,b}$
of different reference masses are related.  It is straightforward to verify
that the transformation laws established in this section under the Lorentz and
the scale transformations are valid for all reference masses.

\begin{center}
{(a)  Lorentz Transformation}
\end{center}

{}From the Lagrange function
\be
{\cal L} = \frac{1}{8\pi} \partial_{\mu}\phi\partial^{\mu}\phi - V(\phi) \ ,
\ee
we can construct the stress tensor as
\be
T_{\mu \nu}  =  \frac{1}{4\pi}\partial_{\mu}\phi\partial_\nu \phi  -  g_{\mu
\nu}{\cal L} \ .
\ee
We assume that ${\cal L}, T_{\mu \nu}$ and $M$ to be introduced are all normal-
ordered w.r.t. mass $m$.  From the definition of $\rho$ and $\overline{\rho}$
in (2.3), we have
\beqn
\phi & = & \rho + \overline{\rho}\\
\dot{\phi} & = & \frac{\partial\rho}{\partial x}  -
\frac{\partial\overline{\rho}}{\partial x} \ .
\eeqn
The Lorentz boost generator is
\beqn
M & = & \int dy(y_0 T_{01} - y_1 T_{00}) \nonumber\\
& = & \int dy\left[t\frac{1}{4\pi} \dot{\phi}\frac{\partial\phi}{\partial y}+
y(\frac{1}{8\pi}\dot{\phi^2}+ \frac{1}{8\pi}(\frac{\partial\phi}{\partial
y})^2 + V(\phi)) \right] \nonumber\\
& = & \frac{1}{4\pi} \int dy\left[(y + t)(\frac{\partial\rho}{\partial y})^2 +
(y - t)(\frac{\partial\overline{\rho}}{\partial y})^2 + 4\pi yV(\rho +
\overline{\rho}) \right] \ .
\eeqn
It is straightforward to verify that $\phi, \frac{\partial\rho}{\partial x}$
and $\frac{\partial\overline{\rho}}{\partial x}$ all transform covariantly
under $M$ with well-defined Lorentz weights $0, \pm 1$.  However, variables
$\rho$ and $\overline{\rho}$ transform noncovariantly as
\bma
\be
[\rho(x,t), M] = i(t\partial_x + x\partial_t) \rho(x,t) - \int dy(x-y)
[\rho(x,t), V(\phi(y,t))]
\ee
\be
[\overline{\rho}(x,t), M] = i(t\partial_x + x\partial_t) \overline{\rho}(x,t)
-\int dy(x-y) [\overline{\rho}(x,t), V(\phi(y,t))] \ .
\ee
\ema
The last potential-dependent terms in (5.6) arise from the nonlocal
definition of $\rho$ and $\overline{\rho}$ (2.3) involving an integration over
space at a fixed time.  Under a Lorentz transformation, the equal-time frame
is changed.  This change induces the nonlocal potential-dependent terms in
(5.6).

The vertex operator $W_{a,b}(x) \equiv D_{a,b}:e^{ia\rho(x) +
ib\overline{\rho}(x)} :$ defined in (2.9) w.r.t a finite mass $m$ transforms
under $M$ as
\be
[W_{a,b}(x), M]  = i[t\partial_x + x\partial_t + \frac{1}{2}(a^2 - b^2)]
W_{a,b}(x) -\int dy(x-y) [W_{a,b}(x) , V(\phi(y))]
\ee
where $\frac{1}{2}(a^2 - b^2)$ represents the Lorentz weight in the absence of
$V(\phi)$.

To verify (5.7), we can follow the same method as described in ref. 4.  The
change of reference mass only leads to an additive constant in $M$, and does
not affect the commutator.

The existence of the extra potential-dependent term in (5.7) implies that the
vertex operator $W_{a,b}(x)$ in general does {\it not} transform covariantly
with a well-defined Lorentz weight.  However, it should be noted that for the
special class of vertex operators $W_{a,b}(x)$ and their adjoint
$W_{-a,-b}(x)$ studied in section 3, defined by
\be
\beta (a-b) = n > 0 ,\qquad  \mbox{$n$ an integer}
\ee
and
\be
\frac{n+1}{2} > a\beta > \frac{n-1}{2},\qquad  \mbox{$n$ = odd}
\ee
\be
\frac{n}{2} + 1 > a\beta > \frac{n}{2} -1, \qquad    \mbox{$n$ = even}\ ,
\ee
the last commutator terms in (5.7) vanish identically.  The cancellation
follows the same argument described in sec. 3.  The extra $(x-y)$ factor only
makes the integration vanish faster.  These special vertex operators not only
satisfy simple equations of motion as shown in sec. 3, they also transform
covariantly with Lorentz weight $\frac{1}{2}(a^2 -b^2)$.

\begin{center}
{(b)  Scale Transformation}
\end{center}

The generator of the scale transformation is
\beqn
S & = & \int dx \ x^{\mu}T_{0\mu}  = \int dx (tT_{00} + xT_{01})\nonumber\\
& = & \frac{1}{4\pi} \int dy \left[(y+t)(\frac{\partial\rho}{\partial y})^2 -
(y-t)(\frac{\partial\overline{\rho}}{\partial y})^2 + 4\pi tV(\rho +
\overline{\rho}) \right] \ .
\eeqn
Unlike the Lorentz generator, $S$ is not a constant of motion.  The time
derivative of $S$ equals to the space integral of $T^{\mu}_{\mu}$, giving
\be
\frac{dS}{dt} = \int dx (T^{\mu}_{\mu}) = 2 \int dy V(\phi(y)) \ .
\ee
In spite of its time dependence, we find it useful to classify field variables
according to their transformation properties under the scale transformation
$S$.  This is analog to describe a system under a nonspherical disturbance by
its angular momentum states.  From (5.11), we can work out the transformation
properties of $\rho$ and $\overline{\rho}$ under $S$ as
\bma
\be
[\rho(x,t), S(t)] = i(t\frac{\partial}{\partial t} +
x\frac{\partial}{\partial x}) \rho(x,t)
\ee
\be
[\overline{\rho}(x,t), S(t)] = i(t\frac{\partial}{\partial t} +
x\frac{\partial}{\partial x}) \overline{\rho}(x,t) \ .
\ee
\ema
Thus, $\rho$ and $\overline{\rho}$ transform covariantly with zero
scaling weight.

We can work out the transformation properties of the vertex operator
$W_{a,b}(x,t)$ in a way similar to those presented in subsection (a).  They
are
\be
[W_{a,b}(x,t), S(t)] = i[t\frac{\partial}{\partial t} +
x\frac{\partial}{\partial x} + \frac{1}{2}(a^2 + b^2)] W_{a,b} (x,t) \ .
\ee
Thus $W_{a,b}(x,t)$ transforms {\it covariantly} under $S(t)$ with a scaling
weight $\frac{1}{2}(a^2 + b^2)$ for all $a$ and $b$, unlike the case of
Lorentz spin.  This is a bit surprising since this non-conformal theory is
Lorentz invariant, but violates scale invariance.  Of course the scale
generator $S(t)$ is not conserved in time.

\begin{center}
{(c) Transformations of Nonlocal Currents}
\end{center}

In this subsection, we shall work out the transformation laws of the currents
$J_{\pm}$ derived in ref. 4, under Lorentz and scale transformations.  To
examine the Lorentz covariance of the currents, we start with
\be
[A_{\frac{2}{\beta}}(x), M] = i\left( t\partial_{x} + x\partial_{t} +
\frac{2}{\beta^{2}} \right) A_{\frac{2}{\beta}}(x) + \int dy (x-y)
[A_{\frac{2}{\beta}}(x) , \lambda \cos(\beta\phi(y))]
\ee
Note that $A_{\frac{2}{\beta}} \equiv W_{\frac{2}{\beta}},0$ corresponds to
the special class of $W_{a,b}$ with $a = \frac{2}{\beta}, b=0, \beta(a-b) =
2$, but is on the boundary of the region $2 \geq \frac{2}{\beta} \geq 0$.
For this case as shown in sec. 3 the commutator $[A_{\frac{2}{\beta}}(x), \cos
\beta\phi(y)]$ does not vanish, but can be evaluated as
\beqn
\lambda \int & dy & (x-y)[A_{\frac{2}{\beta}}(x), \cos\beta\phi(y)] \nonumber\\
& = & \frac{\lambda}{2\overline{m}^2} \int dy(x-y)(-2\pi
i\partial_{x}\delta (x-y)) W_{\frac{2}{\beta}-\beta,-\beta} (y) \nonumber\\
& = & \frac{i\lambda\pi}{\overline{m}^2}
W_{\frac{2}{\beta}-\beta,-\beta} (x) \ .
\eeqn
Hence,
\be
[A_{\frac{2}{\beta}},M] = i(t\partial_{x} + x\partial_{t} +
\frac{2}{\beta^{2}}) A_{\frac{2}{\beta}} +
\frac{i\lambda\pi}{\overline{m}^{2}} W_{\frac{2}{\beta}-\beta,-\beta} \ .
\ee
Similarly, we have
\beqn
[W_{\frac{2}{\beta}-\beta, -\beta}(x),M] & = & i(t\partial_x + x\partial_t +
\frac{2}{\beta^{2}}-2) W_{\frac{2}{\beta}-\beta,-\beta}(x) \nonumber\\
&& -\int dy(x-y) [W_{\frac{2}{\beta}-\beta,-\beta}(x),
\lambda\cos\beta\phi(y)] \ .
\eeqn
For $0 < \beta^{2}< 2$, $W_{\frac{2}{\beta}-\beta,-\beta}$ belongs to the
special class of vertex functions described in (5.10), and the commutator
vanishes.  We have
\be
[W_{\frac{2}{\beta}-\beta},M] = i(t\partial_{x} + x\partial_{t} +
\frac{2}{\beta}-2) W_{\frac{2}{\beta}-\beta,-\beta} \ .
\ee
Eqs. (5.17) and (5.19) imply that $J_{\pm}$ defined in Appendix
A transform covariantly
\be
[J_+, M] = i(t\partial_x + x\partial_t + \frac{2}{\beta^2}) J_+
\ee
\be
[J_-, M] = i(t\partial_x + x\partial_t + \frac{2}{\beta^2}-2) J_- \ .
\ee
Notice that the Lorentz weights of $J_+$ and $J_-$ differ by 2.  Hence the
conservation equation $\partial_- J_+ + \partial_+ J_- = 0$ will also be
Lorentz covariant.

The transformation laws of $A_{\frac{2}{\beta}}$ and
$W_{\frac{2}{\beta}-\beta,-\beta}$ under S can be computed easily as
\be
[A_{\frac{2}{\beta}}(x,t), S(t)] = i(t\frac{\partial}{\partial t} +
x\frac{\partial}{\partial x} + \frac{2}{\beta^2}) A_{\frac{2}{\beta}}(x,t)
\ee
\be
[W_{\frac{2}{\beta}-\beta,-\beta}(x,t), S(t)] = i(t\frac{\partial}{\partial t}
+ x\frac{\partial}{\partial x} + \frac{2}{\beta^2} - 2 + \beta^2)
W_{\frac{2}{\beta}-\beta,-\beta} \ .
\ee
Since these two terms have different scaling weights, as their linear
combination $J_+$ does not have a well-defined scale covariant weight,
although $J_-$ does.  The conservation equation $\partial_- J_+\ +\ \partial_+
J_- = 0$ also is not scale-covariant.

\acknowledgments

R. Rajaraman thanks the Physics Department, University of Illinois at Urbana-
Champaign for their hospitality.  This work was supported in part by the U.S.
National Science Foundation under grant No. 9200148.

\appendix{Renormalization Mass and Infrared Cutoff}

For calculations in the SG theory, we often encounter both the infrared and
the ultraviolet divergences.  To achieve ultraviolet finite results, we
introduce operators in normal-ordered forms, expressed in terms of the
creation and annihilation operators of a free field of mass $m$.  We refer to
this mass $m$ as the normal-ordering reference mass.  In terms of these
normal-ordered operators, the ultraviolet divergences often cancel out.  In
ref. 4, we choose the reference mass to be zero for calculational convenience.
However, this convenience comes with a price.  Most of the calculations with
zero reference mass have infrared divergences.  To achieve finite results, it
is necessary to introduce infrared momentum cutoff $\mu$.  On the other hand,
the calculation with a finite reference mass is more complicated.  One of the
purposes of this appendix is to relate operators defined with respect to
different reference masses, including zero reference mass.  The transformation
properties among general vertex functions involve subtleties which do not
appear to be well known.  We shall also work out explicitly current
conservation laws completely in the framework of a finite reference mass.

The interaction term in the SG theory is
\be
V(\phi) = -\lambda_0 (:\cos\beta\phi:)_0 = -\lambda_m (:\cos\beta\phi:)_m
\ee
where subscripts $0$ and $m$ indicate the normal-ordering reference masses.
The normal-ordered operators and the coupling constants are related by
\be
(:\cos\beta\phi :)_m = \left(\frac{2\mu}{\mu + \sqrt{m^2 + \mu^2}}
\right)^{\beta^2} (:\cos\beta\phi :)_0 \ ,
\ee
\be
\lambda_m = \left(\frac{\mu + \sqrt{m^2 + \mu^2}}{2\mu}
\right)^{\beta^2} \lambda_0 \ .
\ee
For operators and coupling constants associated with two different masses $m,
m' >> \mu$, we have
\be
(:\cos\beta\phi:)_{m'} =
\left(\frac{m}{m'}\right)^{\beta^2}(:\cos\beta\phi:)_m \ ,
\ee
\be
\lambda_m' = (\frac{m'}{m})^{\beta^2}\lambda_m \ .
\ee
Eqs. (A.4) and (A.5) contain neither ultraviolet nor infrared cutoff
dependence.

Turning to general vertex functions $\exp(ia\rho(x) + ib\overline{\rho}(x))$,
we define the normal ordered operator as
\be
(:\exp (ia\rho(x) + ib\overline{\rho}(x)):)_m  =
e^{ia\rho_- + ib\overline{\rho}_-} \ e^{ia\rho_+ + ib\overline{\rho}_+}
\ee
where $\rho_{\pm}$ and $\overline{\rho}_{\pm}$ are related to creation and
annihilation operators by
\bma
\be
\rho_+(x,t) = \int \frac{dk}{\sqrt{\omega}} a(k,t)\ \frac{1}{2}
\ (1-\frac{\omega}{k}) e^{ikx}
\ee
\be
\rho_-(x,t) = \int \frac{dk}{\sqrt{\omega}} a^+(k,t)\  \frac{1}{2}
\ (1-\frac{\omega}{k}) e^{-ikx}
\ee
\be
\overline{\rho}_+(x,t) = \int \frac{dk}{\sqrt{\omega}} a(k,t)\  \frac{1}{2}
\ (1+\frac{\omega}{k}) e^{ikx}
\ee
\be
\overline{\rho}_-(x,t) = \int \frac{dk}{\sqrt{\omega}}\  a^+(k,t)\  \frac{1}{2}
\ (1+\frac{\omega}{k}) e^{-ikx}
\ee
\ema
with $\omega \equiv \sqrt{k^2 + m^2}$.  The exponential operator and the
normal-ordered operator are related by
\be
\exp[ia\rho(x) + ib\overline{\rho}(x)] = (:e^{ia\rho(x) +
ib\overline{\rho}(x)}:)_m \exp[-\frac{1}{2} \langle (a\rho +
b\overline{\rho})^2 \rangle] \ .
\ee
The vacuum expectation values are
\bma
\beqn
\langle \rho\rho \rangle & = & \langle \rho_+(x)\rho_-(x)\rangle
\nonumber\\
& = & \int \frac{dk}{\omega} \theta(\mid k\mid - \mu)\  \frac{1}{4}\
(1-\frac{\omega}{k})^2 \nonumber\\
& = & \ln \frac{2\Lambda}{\mu + \sqrt{m^2 + \mu^2}} + \frac{1}{2} \left(
\frac{\sqrt{m^2 + \mu^2}}{\mu} -1\right),\\
\langle\overline{\rho}\,\overline{\rho}\rangle & = & \int \frac{dk}{\omega}
\theta(\mid k \mid - \mu)\ \frac{1}{4}\ (1 + \frac{\omega}{k})^2 \nonumber\\
& = & \ln\frac{2\Lambda}{\mu + \sqrt{m^2 + \mu^2}} + \frac{1}{2} \left(
\frac{\sqrt{m^2 + \mu^2}}{\mu} -1\right),
\eeqn
and
\beqn
\langle\rho\overline{\rho}\rangle  & = & \int \frac{dk}{\omega}
\theta(\mid k \mid - \mu)\ \frac{1}{4}\ (1 - \frac{\omega^2}{k^2})\nonumber\\
& = & \frac{1}{2} \left(1- \frac{\sqrt{m^2 + \mu^2}}{\mu}\right)
\eeqn
\ema
where we have introduced an ultraviolet cutoff $\Lambda$ and an infrared
cutoff $\mu$.  Substituting (A9) into (A8), we obtain
\beqn
\exp(ia\rho(x) & + & ib\overline{\rho}(x)) = (:e^{ia\rho(x) +
ib\overline{\rho}(x)}:)_m \nonumber\\
& \times & \exp \left[-\frac{a^2 + b^2}{2} \ln\frac{2\Lambda}{\mu + \sqrt{m^2
+ \mu^2}} - \frac{(a-b)^2}{4} (\frac{\sqrt{m^2 + \mu^2}}{\mu} - 1)\right] \ .
\eeqn
For $a \neq b$, there is a strong infrared cutoff dependent factor in the
exponent.  The normal-ordered vertex functions of a finite and a zero
reference mass are related by
\beqn
\left(:e^{ia\rho(x) + ib\overline{\rho}(x)}:\right)_m  & = &
\left(\frac{2\mu}{\mu + \sqrt{\mu^2 + m^2}}\right)^\frac{a^2 + b^2}{2}
\nonumber\\
& \times & \exp\left[\frac{(a-b)^2}{4} (\frac{\sqrt{m^2 + \mu^2}}{\mu} -
1)\right] \left(:e^{ia\rho(x) +ib\overline{\rho}(x)}:\right)_0 \ .
\eeqn
Vertex operators with two different reference masses are related by
\beqn
\left(:e^{ia\rho(x) + ib\overline{\rho}(x)} : \right)_{m'} & = & \left(
\frac{\mu+\sqrt{\mu^2 + m^2}}{\mu + \sqrt{\mu^2 + m'^2}}\right)^
{\frac{a^2 + b^2}{2}} \nonumber\\
& \times & \exp\left[\frac{(a-b)^2}{4\mu} (\sqrt{\mu^2 + m^2}-\sqrt{\mu^2 +
m'^2})\right] \left(:e^{ia\rho(x) + ib\overline{\rho}(x)}:\right)_m
\eeqn
For $a \neq b$, (A12) contains an infrared dependent factor $\exp[(a-b)^2
(m-m')/4\mu]$ as $\mu \rightarrow 0$ which implies that $(:e^{ia\rho +
ib\overline{\rho}}:)_m$ defined above is {\it not} free from infrared
divergence.  A possible candidate for a $\mu$-independent vertex operator is
\beqn
(W_{a,b}(x))_m & = & D_{a,b}(:e^{ia\rho(x) + ib\overline{\rho}(x)}:)_m
\nonumber\\
& \equiv & \exp\left[-(a-b)^2 (\sqrt{\mu^2 + m^2} - \mu) /4\mu\right]
(:e^{ia\rho + ib\overline{\rho}}:)_m \nonumber\\
& = & \left(\frac{2\mu}{\mu + \sqrt{\mu^2 + m^2}}\right)^{\frac{a^2 +b^2}{2}}
(:e^{ia\rho + ib\overline{\rho}}:)_0 \ .
\eeqn
These new normal-ordered operators have infrared independent transformation
law at $m, m' >> \mu$
\be
(W_{a,b}(x))_{m'} = (\frac{m}{m'})^{a^2 + b^2} (W_{a,b}(x))_m \ .
\ee
These $(W_{a,b}(x))_m$ provide $\mu$-independent components for the
construction of conserved currents in the framework of a finite reference
mass.  Note that $(W_{a,b}(x))_0 = (:e^{ia\rho + ib\overline{\rho}}:)_0$ is
the vertex operator defined w.r.t. zero mass, as used extensively in ref. 4.

In ref.4, we have shown that the following currents are conserved $(k_0 = \mu
e^{\gamma})$,
\beqn
(J_+ (x,t))_0 & = & (A_{\frac{2}{\beta}}(x,t))_0 + \frac{\pi\lambda_0}{2\mu^2
e^{2\gamma}}(W_{\frac{2}{\beta}-\beta,-\beta}(x,t))_0 \ ,\\
(J_- (x,t))_0 & = & \frac{\pi\lambda_0}{2\mu^2 e^{2\gamma}}\frac{\beta^2}{2-
\beta^2} (W_{\frac{2}{\beta}-\beta,-\beta}(x,t))_0 \ ,
\eeqn
\be
\partial_- (J_+)_0  + \partial_+ (J_-)_0  =  0 \ .
\ee
In terms of $(W_{a,b}(x))_m$, we introduce
\beqn
(J_+ (x,t))_m & = & (A_{\frac{2}{\beta}} (x,t))_m + \frac{2\pi\lambda_m}{(\mu
+ \sqrt{\mu^2 + m^2})^2e^{2\gamma}} (W_{\frac{2}{\beta}-\beta,-\beta}(x,t))_m
\nonumber\\
& = & \left(\frac{2\mu}{\mu + \sqrt{\mu^2 + m^2}}\right)^{\frac{2}{\beta}}
(J_+(x,t))_0 \\
(J_- (x,t))_m & = & \frac{2\pi\lambda_m}{(\mu + \sqrt{\mu^2 +
m^2})^2e^{2\gamma}} \ \frac{\beta^2}{2-\beta^2}\  (W_{\frac{2}{\beta}-\beta,-
\beta}(x,t))_m \nonumber\\
& = & \left(\frac{2\mu}{\mu + \sqrt{\mu^2 + m^2}}\right)^{\frac{2}{\beta}}
(J_+(x,t))_0
\eeqn
and obtain
\be
\partial_- (J_+)_m + \partial_+ (J_-)_m  = 0.
\ee
Note that the infrared factor $\exp[-(a-b)^2 (\sqrt{m^2 + \mu^2}-\mu)/4\mu]$ is
x-independent, and is the same for all terms of $J_+$ and $J_-$.  Hence,
(A20) follows from (A17).

In the following, we shall derive the above current conservation entirely in
the framework of a finite reference mass $m$.  The equal-time commutation
relations among $\rho$ and $\overline{\rho}$ are still given by (2.5),
independent of the interaction and the reference mass.  However, the
separation of $\rho$ into $\rho_-$ and $\rho_+$ depends on the reference mass,
and so are the equal-time commutation relations among $\rho_{\pm}$ and
$\overline{\rho}_{\pm}$.  They are
\bma
\beqn
[\rho_+ (x,t), \rho_- (y,t)] & = & \int \frac{dk}{\omega} \frac{1}{4}
(1-\frac{\omega}{k})^2 e^{ik(x-y)} \nonumber\\
& = & K_0(m\mid x-y\mid) + \frac{1}{2}M(m\mid x-y\mid) - \frac{i\pi}{2}
\epsilon(x-y)
\eeqn
\be
[\overline{\rho}_+ (x,t), \overline{\rho}_- (y,t)] = K_0(m\mid x-y\mid) +
\frac{1}{2}M(m\mid x-y\mid) + \frac{i\pi}{2} \epsilon(x-y)
\ee
\beqn
[\rho_+ (x,t), \overline{\rho}_- (y,t)] & = &\int \frac{dk}{\omega} \frac{1}{4}
(1-\frac{\omega^2}{k^2}) e^{ik(x-y)} \nonumber\\
& = & -\frac{1}{2} M(m(x-y)) -\frac{i\pi}{2}
\eeqn
\be
[{\rho}_- (x,t), \overline{\rho}_+ (y,t)] = \frac{1}{2} M(m(x-y))
-\frac{i\pi}{2}
\ee
\ema
where $K_0(z)$ is a modified Bessel function, and $M(m(x-y))$ is defined as
\be
M(m(x-y)) \equiv \frac{1}{2} \int \frac{dk}{\omega} \frac{m^2}{k^2}
e^{ik(x-y)}
\ee
and
\be
\frac{d^2M(z)}{dz^2} = -K_0(z) \ .
\ee
The additional $-i\pi/2$ terms in (A.21c) and (A.21d) come from $k=0$ mode as
required by the nonsymmatric definition of $\rho$ and $\overline{\rho}$
relative to $x = \pm\infty$.  $M(m(x-y))$ contains an additive infrared cutoff
dependent constant, $C(m/\mu) = (\sqrt{m^2 + \mu^2} -\mu) /\mu$.  This
infrared dependent constant is the origin of the required infrared factor in
$(W_{a,b})_m$.

In our calculation, we only need commutators between $\rho_{\pm},
\overline{\rho}_{\pm}$ and $\phi_{\pm}$.  These commutators are free of
infrared divergences,
\bma
\beqn
[\rho_+ (x,t), \phi_- (y,t)] & = & \int \frac{dk}{\omega} \frac{1}{2}
(1-\frac{\omega}{k}) e^{ik(x-y)}\nonumber\\
& = & K_0(m\mid x-y\mid) - i\pi\theta(x-y)
\eeqn
\be
[\rho_- (x,t), \phi_+ (y,t)] = -K_0(m\mid x-y\mid) - i\pi\theta(x-y)
\ee
\be
[\overline{\rho}_+ (x,t), \phi_- (y,t)] = K_0(m\mid x-y\mid) +
i\pi\theta(x-y)
\ee
\be
[\overline{\rho}_- (x,t), \phi_+ (y,t)]  = -K_0(m\mid x-y\mid) +
i\pi\theta(x-y) \ .
\ee
\ema
To derive the conservation law, we follow the method described in sec. 3.  The
Heisenberg eq. gives
\beqn
\partial_-(W_{\frac{2}{\beta},0}(x,t))_m  & = & -\frac{i}{2}
\left[(W_{\frac{2}{\beta},0}(x,t))_m, \int dyV(\phi(y))\right]\nonumber\\
& = & i\frac{\lambda_m}{4} \int dy \quad \left[
(W_{\frac{2}{\beta},0}(x))_m,(W_{\beta,\beta}(y) + W_{\-\beta,-
\beta}(y))_m\right] \nonumber\\
& = & \frac{\lambda_m}{4} \int dy \left\{ D_{\frac{2}{\beta},0}
(:e^{\frac{2i}{\beta} \rho(x,t) + i\beta\phi(y,t)}:)_m \right.\nonumber\\
& \times & \left. \left[e^{-2[\rho_+(x,t), \phi_-(y,t)]} -e^{2[\rho_-(x,t),
\phi_+(y,t)]}\right] + (\beta \leftrightarrow -\beta) \right\}\nonumber\\
& = & \frac{\lambda_m}{4} \int dy \left\{ D_{\frac{2}{\beta},0}
(:e^{\frac{2i}{\beta} \rho(x,t) + i\beta\phi(y,t)}:)_m \right.\\
& \times & \left. \left[ e^{-2K_0(m\mid x-y\mid) + 2i\pi\theta(x-y)} -e^{-
2K_0(m\mid x-y\mid) - 2i\pi\theta(x-y)}\right] + (\beta \leftrightarrow -
\beta) \right\}. \nonumber
\eeqn
For $x \neq y$, the integrand vanishes.  Thus, all the contribution comes from
the infinitesimal region around $y = x$.  Near $y = x$, we have
\be
K_0(m\mid x-y\mid) = -\gamma -\ln(\frac{1}{2} m\mid x-y\mid) + 0((x-y)^2)
\ee
\beqn
[\overline{\rho}_- (x,t), \phi_+ (y,t)] & = & \gamma +\ln(\frac{1}{2} m\mid
x-y\mid) -i\pi\theta(x-y) + 0((x-y)^2)\nonumber\\
& = & \ln(-i\overline{m}(x-y + i\epsilon)) -\frac{i\pi}{2} + 0((x-y)^2)
\eeqn
where
\be
\overline{m} \equiv \frac{1}{2}me^2 \ .
\ee
Eq. (A.27) has the same short distance structure as those obtained from (2.7)
- (2.9) of ref. 4 with the replacement of $k_0$ by $\overline{m}$. This is in
fact true for other terms in the commutators, and the derivation of new
currents in ref. 4 follow through.  This leads to the construction of a
conserved current which has the same structure of (A.15) (A.16), ((3.11) and
(3.12) of ref. 4) with the replacement of $(W_{a,b})_0$ by $(W_{a,b})_m$,
$\lambda_0$ by $\lambda_m$, and $k_0$ by $\overline{m}$, giving (at $m$ finite
and $\mu \rightarrow 0$),
\beqn
(J_+(x,t))_m & = & (W_{\frac{2}{\beta},0} (x,t))_m + \frac{2\pi\lambda_
m}{m^2 e^{2\gamma}} (W_{\frac{2}{\beta}-\beta,-\beta}(x,t))_m \\
(J_-(x,t))_m & = & \frac{2\pi\lambda_m}{m^2 e^{2\gamma}}\  \frac{\beta^2}
{2-\beta^2} (W_{\frac{2}{\beta}-\beta,-\beta}(x,t))_m
\eeqn
as desired.

\appendix{Contributions From Short Distance Singularies}

Consider the integral
\be
I = \lim_{\epsilon\rightarrow 0} \int_{-L}^L dx(x^2 + \epsilon^2)^p ((x +
i\epsilon)^n - (x - i\epsilon)^n) C(x)
\ee
where $n$ is a positive integer, $p$ is real and $C(x)$ is a nonsingular
function.  We see that the integrand vanishes as $\epsilon \rightarrow 0$ for
any $x \neq 0$.  Hence the integral will receive nonzero contributions, if
any, only from possible singularities at $x = 0$.  We restrict the range of
$x$ to be in $[-L, L]$ with $L^{-1} \sim \mu$, the infrared mass we have used
throughout the paper. We evaluate this integral with the understanding that
$\epsilon \rightarrow 0$ first, and then $L$ can tend to infinity.  Expand
\[
C(x) = \sum_{m=0}^{\infty} x^m C_m \ ,
\]
\[
(x + i\epsilon)^n = \sum_{r=0}^n x^{n-r}(i\epsilon)^r \frac{n!}{r!(n-r)!} \ ,
\]
and change variables to $y=x/\epsilon$.  Then,
\be
I = \lim_{\epsilon\rightarrow 0} \ \sum_{m=0}^{\infty} \ \sum_{odd\ r}^n
\ \epsilon^{2p + 1 + n + m} K_{mr}
\int^{L/\epsilon}_{-L/\epsilon} dy(1+y^2)^{2p} y^{n + m -r}
\ee
where
\be
K_{mr} = \frac{2(i)^rn!}{r!(n-r)!} C_m \ .
\ee
As $\epsilon\rightarrow 0$, the limits of the y-integration tend towards
infinity.  For large $y$, the y-integrand behaves as $y^{2p +n+m-r}$. Let $N$
be the largest (positive or negative) integer less than $(-2p-n-1)$. Then for
$m-r \leq N$ the y-integral converges even when $\epsilon\rightarrow 0$,
whereas, for $m-r > N$, it behaves as $(L/\epsilon)^{2p+n+m-r+1}$.  Hence $I$
has the form
\be
I = \lim_{\epsilon\rightarrow 0} \left[\sum^n_{odd\ r} \sum^{m-r\leq N}_m
\epsilon^{2p + n + m + 1} A_{rm}  + \sum^{m-r>N}_m \sum^n_{odd\ r}
\epsilon^r B_{rm} \right]
\ee
where $A_{rm}$ and $B_{rm}$ represent the result of y-integrals multiplied by
finite factors like $K_{rm}$ in (B.3).  For any finite $L$, as
$\epsilon\rightarrow 0$, the second term vanishes since $r$ is positive.
Turning to the first term, notice that the y-integrals in (B.2) vanish by $y
\leftrightarrow -y$ symmetry unless $n+m-r$ is even.  Since $r$ is odd anyway,
this requires $n+m$ odd.  Thus, for even (odd) $n$, the lowest value of $m$
that contributed is one (zero).  Hence we conclude, for different cases:

\noindent (a) $n$ odd
\bma
\beqn
  (i) \quad & \mbox{when}\ 2p + n + 1 > 0  \qquad & I = 0\\
 (ii) \quad & \mbox{when}\ 2p + n + 1 = 0  \qquad & I = \mbox{finite}\\
(iii) \quad & \mbox{when}\ 2p + n + 1 < 0  \qquad & I \mbox{ diverges as
                                                 $\epsilon\rightarrow 0$}\ .
\eeqn
\ema
Notice that in cases (ii) and (iii) above, there will exist some finite
nonzero $A_{rm}$, satisfying $m-r\leq N$, provided the function $C(x)$ and its
first derivative are nonzero at $x=0$.

\noindent (b) $n$ even
\bma
\beqn
  (i) \quad & \mbox{when}\ 2p + n + 2 > 0 \qquad & I = 0\\
 (ii) \quad & \mbox{when}\ 2p + n + 2 = 0 \qquad & I = \mbox{finite}\\
(iii) \quad & \mbox{when}\ 2p + n + 2 < 0 \qquad & I \mbox{ diverges as
                                               $\epsilon\rightarrow 0$} \ .
\eeqn
\ema

\end{document}